\documentclass[twocolumn,amsmath,showpacs,amsfonts,aps,prc,floatfix]{revtex4}
\usepackage{graphicx}
\usepackage{bm}
\newcommand{\dn}{\frac{dN^\phi}{dy}} 
\newcommand{\pt}{<p_T^\phi>} 
\begin{document}

\title{Initial eccentricity and constituent quark number scaling of elliptic flow in ideal and viscous dynamics} 
 
\author{A. K. Chaudhuri}
\email[E-mail:]{akc@veccal.ernet.in}
\affiliation{Variable Energy Cyclotron Centre, 1/AF, Bidhan Nagar, 
Kolkata 700~064, India}

\begin{abstract}

In the Israel-Stewart's theory of dissipative hydrodynamics,   
we study the scaling properties   elliptic flow in Au+Au collisions.
Initial energy density of the fluid was fixed to reproduce STAR data on $\phi$ meson multiplicity in 0-5\% Au+Au collisions, such that irrespective of fluid viscosity, entropy at the freeze-out is similar in ideal or in viscous evolution.
Initial eccentricity  or constituent quark number scaling is only approximate in ideal or minimally viscous ($\eta/s=1/4\pi$) fluid. Eccentricity scaling become nearly exact in more viscous fluid ($\eta/s \geq$0.12). However, in more viscous fluid,   constituent quark number scaled elliptic flow for mesons and baryons split into separate scaling functions. Simulated flows also do not exhibit 'universal scaling' i.e. elliptic flow scaled by the constituent quark number and charged particles $v_2$ is not a single function of transverse kinetic energy scaled by the quark number. From a study of violation of universal scaling, we obtain an estimate of QGP viscosity, $\eta/s=0.12 \pm 0.03$. The error is statistical only. Systematic error  in $\eta/s$ could be as large.

  \end{abstract}

\pacs{47.75.+f, 25.75.-q, 25.75.Ld} 

\date{\today}  

\maketitle


\section{Introduction} \label{intro}
 
Recently STAR collaboration published their analysis of elliptic flow ($v_2$) of identified particles in Au+Au collisions at $\sqrt{s}$=200 GeV \cite{:2008ed}. In the relatively low $p_T$ region, $p_T <$ 2GeV, $v_2$ of identified particles, in each centrality bins studied, scales
with transverse kinetic energy $KE_T=m_T-m$. However, they did not find initial
or participant  eccentricity scaling of $v_2(p_T)$ or $v_2(m_T-m)$.
At higher $p_T$, $v_2$ scales with constituent quark number.  They also observed
'universal scaling', $v_2/(n_q <v_2>_{ch})$, elliptic flow scaled by the constituent quark numbers and charged particles $v_2$, of different particle species in different collision centrality, scales with $KE_T/n_q$, the transverse kinetic energy per constituent quark number.  Similar results are obtained in PHENIX measurements \cite{Taranenko:2007gb}.  
The constituent quark number scaling of elliptic flow  suggests existence of an initial collective partonic state. Constituent quark number scaling of elliptic flow is naturally explained in a coalescence model   \cite{Molnar:2003ff,Scheibl:1998tk}, where elliptic flow of constituent quarks adds up. 

Elliptic flow in heavy ion collisions is best understood in a hydrodynamic model \cite{QGP3}. Elliptic flow measure the momentum anisotropy.  In non-zero impact parameter collisions, the reaction zone is spatially asymmetric. Differential pressure gradient convert the spatial asymmetry to momentum asymmetry. Approximate initial eccentricity scaling of elliptic flow is observed in ideal hydrodynamic model
simulations for Au+Au  collisions. Constituent quark number scaling, however is not indicated in ideal hydrodynamic simulation of Au+Au collisions. 
In the present paper, we investigate the scaling properties of elliptic flow in ideal and viscous hydrodynamics. From the study of scaling violation, we   also    obtain a phenomenological estimate of QGP viscosity, $\eta/s=0.12 \pm 0.03$.
In recent years, there is considerable interest in viscosity of  strongly interacting Quark-Gluon Plasma.
Theoretical estimate cover a wide range, $\eta/s\approx$ 0.0-1.0 . 
String theory based models (ADS/CFT) give a lower bound on viscosity of any matter $\eta/s \geq 1/4\pi$ \cite{Policastro:2001yc}. In a perturbative QCD, Arnold et al  \cite{Arnold:2000dr} estimated $\eta/s\sim$ 1.  In a SU(3) gauge theory, Meyer \cite{Meyer:2007ic} gave the upper bound $\eta/s <$1.0, and his best estimate is $\eta/s$=0.134(33) at $T=1.165T_c$.  At RHIC region, Nakamura and Sakai \cite{Nakamura:2005yf} estimated the viscosity of a hot gluon gas  as $\eta/s$=0.1-0.4. Attempts have been made to estimate QGP viscosity directly from experimental data.  Gavin and Abdel-Aziz \cite{Gavin:2006xd} proposed to measure viscosity from transverse momentum fluctuations. From the existing data on Au+Au collisions, they estimated that QGP viscosity as $\eta/s$=0.08-0.30. Experimental data on elliptic flow has also been used to estimate QGP viscosity. Elliptic flow scales with eccentricity. Departure from the scaling can be understood as due to off-equilibrium effect and utilized to estimate viscosity \cite{Drescher:2007cd} as, $\eta/s$=0.11-0.19. Experimental observation that elliptic flow scales with transverse kinetic energy is also used to estimate QGP viscosity, $\eta/s \sim$ 0.09 $\pm$ 0.015 \cite{Lacey:2006bc}, a value close to the ADS/CFT bound. From heavy quark energy loss, PHENIX collaboration \cite{Adare:2006nq} estimated QGP viscosity $\eta/s\approx$ 0.1-0.16.  In \cite{Luzum:2008cw,Song:2008hj},
 from analysis of RHIC data, an upper bound to viscosity is given $\eta/s <$ 0.5 \cite{Luzum:2008cw,Song:2008hj}. Recently 
in   \cite{Chaudhuri:2009uk}, from a detail analysis of centrality dependence of $\phi$ meson multiplicity, mean $p_T$ and integrated $v_2$, QGP viscosity was estimated as, $\eta/s=0.07 \pm 0.03 \pm 0.14$, the first error is statistical, the second one is systematic. The large systematic error is due to uncertain knowledge about various parameters of a hydrodynamics model, e.g. initial time, energy density profile, freeze-out condition,, finite accuracy of computation etc. However, estimates of QGP viscosity from experimental data must be treated with caution. Recently, in \cite{Chaudhuri:2009hj}, it was shown that viscosity to entropy ratio ($\eta/s$) in Au+Au collisions depend on the collision centrality.  While in central collisions, charged particle elliptic flow demand nearly perfect fluid, more viscous fluid is demanded in peripheral collisions. In the collision centrality, 0-60\%, $\eta/s$ can vary from 0-0.17. 

The paper is organized as follows: in section.\ref{sec2}, we describe the model: hydrodynamic equations for space-time evolution of the fluid, equation of state and initial conditions. Various    scaling properties of elliptic flow in ideal and viscous fluid evolution are studied in section.\ref{sec3}. In section.\ref{sec4}, from violation of universal scaling, we obtain an estimate of viscosity to entropy ratio. Results are summarized in section.\ref{sec5}. 

\section{Hydrodynamic equations, equation of state and initial conditions} \label{sec2}

Israel-Stewart's theory of 2nd order dissipative relativistic hydrodynamics is well established \cite{IS79}.
Briefly, in Israel-Stewart's theory, the thermodynamic space is extended to include the dissipative flows, relaxation equations for which are solved simultaneously  with the energy-momentum, baryon number  conservation equations. More detail exposition can be found in \cite{Heinz:2005bw,Chaudhuri:2008sj}. 

Presently, we assume that in $\sqrt{s}_{NN}$=200 GeV, Au+Au collisions  at RHIC, a baryon free fluid is formed.
Only dissipative effect we consider is the shear viscosity, heat conduction and bulk viscosity is neglected.  
The space-time evolution of the fluid is obtained by solving,

\begin{eqnarray}  
\partial_\mu T^{\mu\nu} & = & 0,  \label{eq3} \\
D\pi^{\mu\nu} & = & -\frac{1}{\tau_\pi} (\pi^{\mu\nu}-2\eta \nabla^{<\mu} u^{\nu>}) \nonumber \\
&-&[u^\mu\pi^{\nu\lambda}+u^\nu\pi^{\nu\lambda}]Du_\lambda. \label{eq4}
\end{eqnarray}

Eq.\ref{eq3} is the conservation equation for the energy-momentum tensor, $T^{\mu\nu}=(\varepsilon+p)u^\mu u^\nu - pg^{\mu\nu}+\pi^{\mu\nu}$, 
$\varepsilon$, $p$ and $u$ being the energy density, pressure and fluid velocity respectively. $\pi^{\mu\nu}$ is the shear stress tensor. Eq.\ref{eq4} is the relaxation equation for the shear stress tensor $\pi^{\mu\nu}$.   
In Eq.\ref{eq4}, $D=u^\mu \partial_\mu$ is the convective time derivative, $\nabla^{<\mu} u^{\nu>}= \frac{1}{2}(\nabla^\mu u^\nu + \nabla^\nu u^\mu)-\frac{1}{3}  
(\partial . u) (g^{\mu\nu}-u^\mu u^\nu)$ is a symmetric traceless tensor. $\eta$ is the shear viscosity and $\tau_\pi$ is the relaxation time.  It may be mentioned that in a conformally symmetric fluid relaxation equation can contain additional terms  \cite{Song:2008si}.

Assuming boost-invariance, Eqs.\ref{eq3} and \ref{eq4}  are solved in $(\tau=\sqrt{t^2-z^2},x,y,\eta_s=\frac{1}{2}\ln\frac{t+z}{t-z})$ coordinates, with the code 
  "`AZHYDRO-KOLKATA"', developed at the Cyclotron Centre, Kolkata.
 Details of the code can be found in \cite{Chaudhuri:2008sj}. 
Within 10\% or less, AZHYDRO-KOLKATA simulation  reproduces  Song and Heinz's  \cite{Song:2008si} result for temporal evolution of momentum anisotropy $\varepsilon_p$.

 \begin{figure}[t]
\center
 \resizebox{0.35\textwidth}{!}{%
  \includegraphics{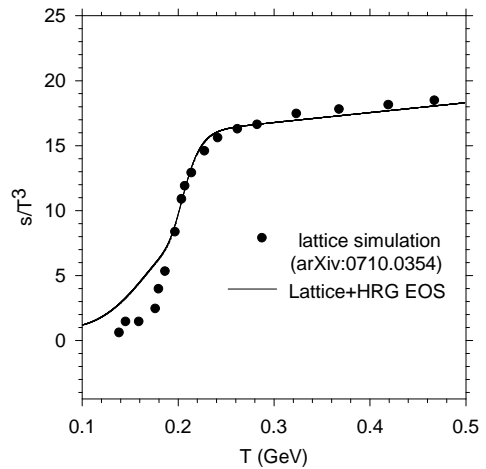}
}
\caption{Black circles are lattice simulation \cite{Cheng:2007jq} for entropy density. The black line is the model EOS, obtained by parametric representation to the lattice simulations and a hadronic resonance gas at low temperature.}
  \label{F1}
\end{figure} 

Eqs.\ref{eq3},\ref{eq4} are closed with an equation of state $p=p(\varepsilon)$.
Lattice simulations \cite{lattice,Cheng:2007jq} indicate that the confinement-deconfinement transition is a cross over, rather than a 1st or 2nd order phase transition.   In Fig.\ref{F1},  a recent lattice simulation  \cite{Cheng:2007jq} for the  entropy density is
  shown. We complement the lattice simulated EOS \cite{Cheng:2007jq} by a
  hadronic resonance gas (HRG) EOS comprising all the resonances below mass 2.5 GeV. In Fig.\ref{F1}, the solid line is the
   entropy density of the "`lattice +HRG"' EOS. The entropy density is obtained as,
     
   \begin{equation}
   s=0.5[1-tanh(x)]s_{HRG} + 0.5 [1-tanh(x)]s_{lattice}
   \end{equation}
   
\noindent   with $x=\frac{T-T_c}{\Delta T}$. In the present simulation, we have used cross over temperature, $T_c$=196 MeV and $\Delta T=0.1T_c$. Compared to lattice simulation, entropy density in HRG drops slowly at low temperature.
It is consistent with observation in \cite{Cheng:2007jq}, that at low temperature, trace anomaly,
$\frac{\varepsilon-3p}{T^4}$ drops faster in lattice simulation than in a HRG model. It is difficult to resolve whether the discrepancy is due to failure of HRG model at lower temperature
or due to the difficulty in resolving low energy hadron spectrum on   rather coarse lattice \cite{Cheng:2007jq}.

Solution of partial differential equations (Eqs.\ref{eq3},\ref{eq4}) requires initial conditions, e.g.  transverse profile of the energy density ($\varepsilon(x,y)$), fluid velocity ($v_x(x,y),v_y(x,y)$) and shear stress tensor ($\pi^{\mu\nu}(x,y)$) at the initial time $\tau_i$. One also need to specify the viscosity ($\eta$) and the relaxation time ($\tau_\pi$). A freeze-out prescription is also needed to convert the information about fluid energy density and velocity to particle spectra and compare with experiment.

We initialize the fluid as in \cite{Chaudhuri:2009uk}.
We assume that the fluid is thermalized at $\tau_i$=0.6 fm \cite{QGP3} and the initial fluid velocity is zero, $v_x(x,y)=v_y(x,y)=0$.   Initial energy density is assumed to be distributed as \cite{QGP3}

\begin{equation} \label{eq6}
\varepsilon({\bf b},x,y)=\varepsilon_i[0.75 N_{part}({\bf b},x,y) +0.25 N_{coll}({\bf b},x,y)],
\end{equation}

\begin{figure}[t]
 \center
 \resizebox{0.35\textwidth}{!}{%
  \includegraphics{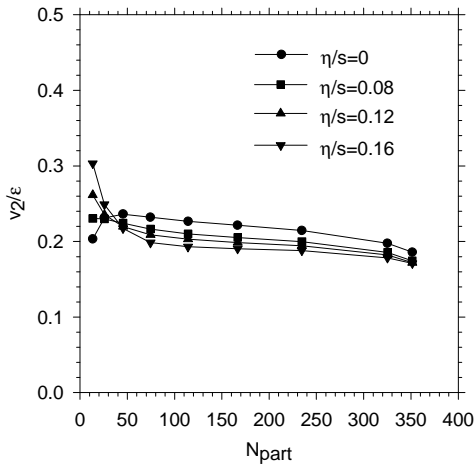} 
}
\caption{charged particle's integrated 
elliptic flow scaled by the participant eccentricity ($v_2/\varepsilon$) 
as a function of collision centrality. The lines with the filled circles, squares, triangles and inverted triangles are hydrodynamic predicted $v_2$ in   fluid evolution with $\eta/s$=0, 0.08, 0.12 and 0.16 respectively.  }
\label{F2}
\end{figure} 

\noindent
where b is the impact parameter of the collision. $N_{part}$ and $N_{coll}$ are the transverse profile of the average number of  participants and average number collisions respectively.
$\varepsilon_i$ is a parameter which does not depend on the impact parameter of the collision.   The shear stress tensor was initialized with boost-invariant value, $\pi^{xx}=\pi^{yy}=2\eta/3\tau_i$, $\pi^{xy}$=0. For the relaxation time, we used the   Boltzmann estimate $\tau_\pi=3\eta/2p$.  
 Hydrodynamics also require a freeze-out condition. In the present work,   we have assumed baryon free fluid. The chemical potential is zero throughout the evolution. We assume that the kinetic freeze-out occur at a fixed temperature $T_F$=150 MeV \cite{note1}. Details of particle production in viscous hydrodynamics can be found in \cite{Chaudhuri:2008sj}. Briefly, invariant particle distribution from the freeze-out  hyper-surface $\Sigma_\mu$   is obtained using Cooper-Frye prescription   \cite{Cooper:1974mv}, 

\begin{equation} \label{eq6_1}
E\frac{dN}{d^3p}= \frac{dN}{dy d^2p_T} =\int_\Sigma d\Sigma_\mu p^\mu f(x,p)
\end{equation}

\noindent where $f(x,p)$ is the
one-body distribution function with a non-equilibrium correction, $f(x,p)=f_{eq}(x,p)[1+\phi(x,p)]$, $\phi(x,p)=\frac{1}{2(\varepsilon+p)T^2} \pi^{\mu\nu} p_\mu p_\nu$.  


\begin{figure}[t]
 \center
 \resizebox{0.35\textwidth}{!}{%
  \includegraphics{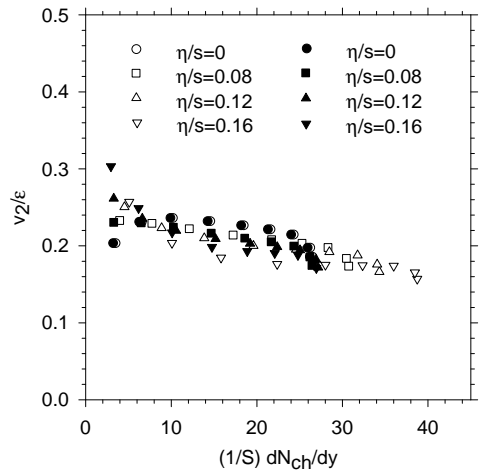} 
}
\caption{charged particle's integrated 
elliptic flow scaled by the participant eccentricity ($v_2/\varepsilon$) 
as a function of $\frac{1}{S}\frac{dN_{ch}}{dy}$, in ideal and viscous evolution.  The blank symbols
are the scaled flow when central energy density is fixed at $\varepsilon_0$=35.5 $GeV/fm^3$. The filled symbols are obtained in evolution when central energy density is reduced to keep multiplicity fixed. For ease of visibility, blank circles are shifted horizontally.}
\label{F3}
\end{figure} 

\begin{table}[ht]
\caption{\label{table1} Initial central energy density ($\varepsilon_i$) and temperature ($T_i$) of the fluid in b=0 Au+Au collisions, for different values of viscosity to entropy ratio ($\eta/s$). The predicted $\phi$ meson multiplicity
and mean $p_T$ are also noted. They should be compared with STAR measurements,
  ${\dn}_{ex}=7.95\pm0.74$ and ${\pt}_{ex}=0.977 \pm 0.064$.
} 
\begin{ruledtabular} 
  \begin{tabular}{|c|c|c|c|c|}\hline
$\eta/s$         & 0    & 0.08 & 0.12 & 0.16 \\  \hline
$\varepsilon_i (GeV/fm^3)$ & $35.5$ & $29.1$ & $25.6$ &  $20.8$ \\  
  & $\pm$ 5.0 & $\pm$ 3.6 & $\pm$ 4.0 &  $\pm$ 2.7  \\ \hline
$T_i$ (MeV) & 377.0 & 359.1 & 348.0 & 330.5     \\ 
  & $\pm 13.7$ & $\pm 11.5$ & $\pm 14.3$ & $\pm 11.3$   \\ \hline  
$\frac{dN^\phi}{dy}$ & 7.96 & 8.01 &  8.22 & 8.13\\ \hline
$<p_T\phi>$ & 1.019 & 1.062 &  1.111 & 1.174\\ \hline
 \end{tabular}\end{ruledtabular}  
\end{table}  

We have simulated Au+Au collisions for four values of viscosity, (i) $\eta/s$=0 (ideal fluid), (ii) $\eta/s=1/4\pi\approx$ 0.08 (ADS/CFT lower limit of viscosity), (iii)   $\eta/s$=0.12 and (iv) $\eta/s$=0.16.
In viscous fluid evolution, entropy is generated. To obtain a common ground to compare between ideal and viscous evolution,
we fix the initial energy density of the fluid such that entropy at the freeze-out, both in ideal and viscous fluid evolution, is the same. This is done by reproducing the STAR data  \cite{Abelev:2007rw}, on $\phi$ meson multiplicity in 0-5\% Au+Au collisions. 
In table 1,   initial central energy density ($\varepsilon_i$) and temperature ($T_i$) required to fit STAR data on $\phi$ meson multiplicity in 0-5\% Au+Au collisions are noted. The error in $\varepsilon_i$ or in $T_i$ corresponds to statistical and systematic uncertainty in STAR measurements \cite{Abelev:2007rw}.
 The predicted $\phi$ meson multiplicities and mean $p_T$  are also shown in table.\ref{table1}. Irrespective of fluid viscosity, initial energy density (or temperature) of the fluid can be tuned to reproduce  
STAR measurements for $\phi$ meson multiplicity, ${\dn}^{ex}=7.95 \pm 0.74$ (statistical and systematic error included). As entropy production increases with viscosity, with increasing viscosity, fluid require less and less initial energy density.
For example, compared to ideal fluid in minimally viscous fluid, initial energy density is reduced by $\sim$ 18\%. In fluid with viscosity $\eta/s$=0.16, reduction is even more, $\sim$40\%. In table.\ref{table1}, we have also noted the hydrodynamic predictions for $\phi$ meson mean $p_T$. $<p_T>$ increases in viscous evolution and one find that  predictions overestimate the STAR measurement, $\pt=0.977\pm 0.064$ (statistical and systematic error included) for $\eta/s \geq$0.12. Experimental data on $\phi$ meson multiplicity and mean $p_T$ in 0-5\% centrality Au+Au collisions are simultaneously explained only for $\eta/s \leq$0.12.

\begin{figure}[t]
 \center
 \resizebox{0.35\textwidth}{!}{%
  \includegraphics{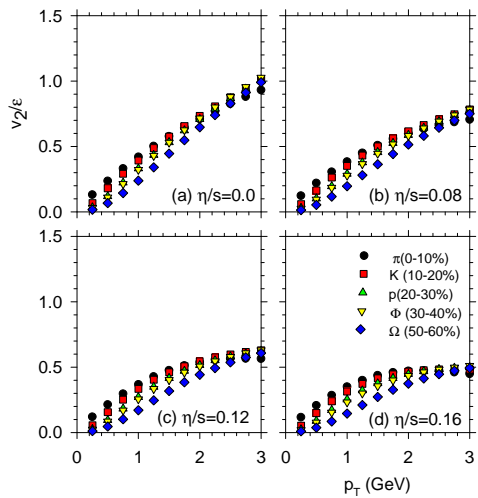} 
}
\caption{(color online) 
Elliptic flow scaled by the participant eccentricity ($v_2/\varepsilon$) 
as a function of transverse momentum. The different symbols are for 
$\pi^-$, $K^+$, $p$, $\phi$ and $\Omega$ in   evolution of QGP fluid with viscosity to entropy ratio (a)$\eta/s$=0, (b)$\eta/s$=0.08, (c) $\eta/s$=0.12 and (d) $\eta/s$=0.16. }\label{F4}
\end{figure} 

\section{Scaling properties of elliptic flow} \label{sec3}

\subsection{Initial eccentricity scaling}

We first investigate the initial eccentricity scaling of integrated elliptic flow in ideal and viscous dynamics. It is now realized that the experimentally measured scaling of integrated $v_2$ with multiplicity or with collision centrality
is not in agreement with hydrodynamics. While hydrodynamics predicts approximate
scaling \cite{Song:2008si}, in experiments scaling is violated \cite{Voloshin:2007af,Voloshin:2006wi}. As discussed in \cite{Bhalerao:2005mm}, violation of the scaling can be understood as an indication of incomplete thermalization.
However, there is a serious flaw in the arguments of \cite{Bhalerao:2005mm}. In 
\cite{Bhalerao:2005mm}, a  
relation between inverse of the Knudsen parameter $\frac{1}{K}$, and multiplicity per unit the transverse area  $\frac{1}{S}\frac{dN}{dy}$,  was obtained under the assumption that the total particle number is conserved throughout the evolution. The assumption was justified with the observation that particle density is proportional to entropy density and entropy is conserved.   In a viscous evolution, entropy is generated and initial and final state entropy are not same and the assumption is clearly violated.  Only in systems with very small viscosity, the assumption may be valid, but not in systems where sufficient entropy is generated. Explicit  numerical simulations show (see table 1) that entropy generation can be substantial, e.g. $\sim$ 20\%, 30\% and 50\% 
in fluid evolution with  viscosity to entropy ratio, $\eta/s$=0.08, 0.12 and 0.16.

 \begin{figure}[t]
 \center
 \resizebox{0.35\textwidth}{!}{%
  \includegraphics{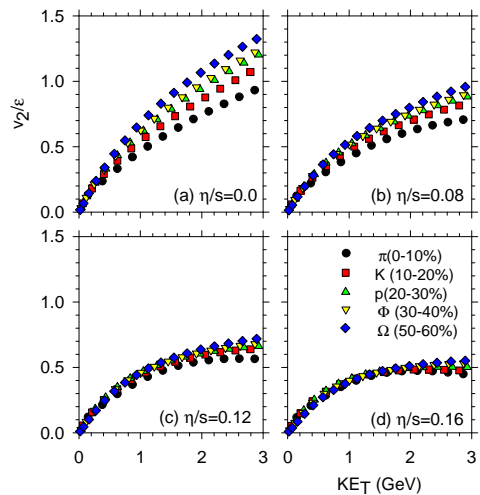} 
}
\caption{(color online) same as in Fig.\ref{F3}, but as a function of transverse kinetic energy.}
 \label{F5}
\end{figure} 

In Fig.\ref{F2}, present model predictions for the eccentricity scaled charged particle's elliptic flow,
in ideal and viscous dynamics, as a function of participant numbers ($N_{part}$) are shown. Except for the very peripheral collisions, $v_2/\varepsilon$,
in ideal or viscous dynamics, approximately scales with collision centrality.
In very peripheral collisions, $N_{part}<$50, $v_2$ in viscous hydrodynamics is more than that in ideal hydrodynamics. However, applicability of hydrodynamics in very peripheral collisions is questionable \cite{QGP3}. Apparently, the results appear to contradict simulation studies of Song and Heinz \cite{Song:2008si}.
In \cite{Song:2008si}, Song and Heinz studied initial eccentricity scaling of integrated $v_2$ as a function of charged particle multiplicity per unit transverse area, $\frac{1}{S}\frac{dN}{dy}$. They obtained     approximate scaling  both in ideal and viscous fluid, however, scaling violation is somewhat larger in viscous dynamics. In \cite{Song:2008si} fluid was not initialized to have similar multiplicity. Viscosity reduces elliptic flow. However, since entropy is generated in viscous evolution, particle multiplicity is increased. 
A part of larger scaling violation in viscous dynamics is due to enhanced multiplicity and reduced elliptic flow. In the present simulation, fluid was initialized to have similar multiplicity, both in ideal and viscous dynamics. 
Since energy density is reduced, viscous suppression is also comparatively less.
Consequently, scaling violation is not large in viscous fluid evolution.  
This is explicitly shown in Fig.\ref{F3}.
In Fig.\ref{F3},  the blank circles, squares, up triangles and down triangles    corresponds to eccentricity scaled elliptic flow in hydrodynamic simulation with viscosity to entropy ratio $\eta/s$=0, 0.08, 0.12 and 0.16 respectively. The initial central energy density is fixed at $\varepsilon_0$=35.5 $GeV/fm^3$. With increasing viscosity,  $\frac{1}{S}\frac{dN}{dy}$ increases but elliptic flow decreases.
The filled symbols in  Fig.\ref{F3} are the scaled flow when multiplicity is constrained by reducing central energy density in more viscous fluid (see table.\ref{table1}).   With the exception of very peripheral collisions,
dispersion between the filled symbols is comparatively less than that between the blank symbols. For a given viscosity, elliptic flow is less suppressed when multiplicity is constrained.

 \begin{figure}[t]
 \center\resizebox{0.35\textwidth}{!}{%
  \includegraphics{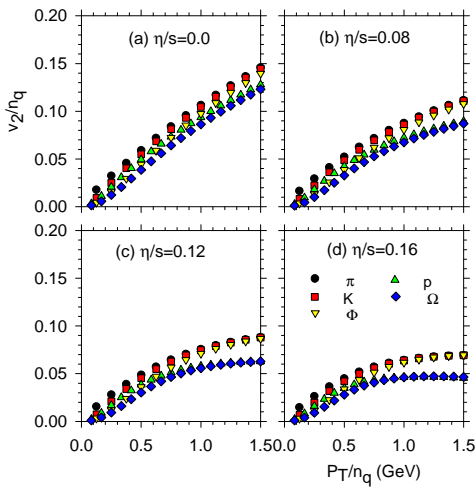} 
}
\caption{(color online) 
Elliptic flow scaled by the 
constituent quark number   for $\pi^-$, $K^+$, $p$, $\phi$ and $\Omega$ in 20-30\% Au+Au collision
as a function of transverse momentum scaled by the constituent quark number ($p_T/n_q$).
Scaled flow in deal and viscous $\eta/s$=0.08, 0.12 and 0.16 fluid are shown   separately.}
\label{F6}
 \end{figure} 
 
To investigate the initial eccentricity scaling of differential elliptic flow of different particle species in different collision centrality, we have computed elliptic flow for $\pi^-$, $K^+$, proton, $\phi$ and $\Omega$ in 0-10\%, 10-20\%, 20-30\%, 30-40\%, 40-50\% and 50-60\% Au+Au collisions. 
Note the particle list contain non-strange and strange mesons and baryons. In a hydrodynamic model, 
  elliptic flow  
of identified particles, in different collision centrality,  approximately scales. To show the scaling between different particle species, in Fig.\ref{F4}, scaled elliptic flow for (i) $\pi^-$ in 0-10\%, (ii) $K^+$ in 10-20\%, (iii) $\phi$ in 20-30\%,
(iv) $p$ in 30-40\% and (vi) $\Omega$ in 50-60\% Au+Au collisions are shown.   Only a few collision centralities  are chosen to show the scaling behavior clearly. Results are essentially same if some other centralities are chosen.  
In the present work, we have neglected resonance decays. Resonance contribution reduces elliptic flow, mostly at low $p_T$ \cite{Hirano:2002ds}. Pions are most affected by resonance decays. In the low $p_T$ range, $0 \leq p_T \leq 1$ GeV, $v_2$ is reduced by $\sim$ 0-30\%. At $p_T >$ 1 GeV, resonance contribution to $v_2$ is negligible \cite{Hirano:2002ds}. 
Mass ordering of the flow is clearly evident at low $p_T$, more $v_2$ for lighter particles. But at large $p_T$ mass effect is reduced.  One observe that as a function of 
$p_T$ elliptic flow do not show initial eccentricity scaling. Same results are shown in Fig.\ref{F5} as a function of transverse kinetic energy, $KE_T=\sqrt{m^2+p_T^2}-m$. As a function of
 $KE_T$, initial eccentricity scaling of elliptic flow is only approximate in ideal fluid evolution. Scaling gets better with viscosity.  
 For example, in ideal hydrodynamics, at $KE_T\approx$ 2 GeV,  $v_2/\varepsilon$ for $\Omega$ in 50-60\% Au+Au collisions is $\sim$45\% larger than the same for $\pi^-$ in 0-10\% Au+Au collisions.
 The difference is less in viscous evolution, e.g. $\Omega$ flow exceed that of $\pi$ by,  $\sim$ 30\%, $\sim$15\% and $\sim$7\% for $\eta/s$=0.08, 0.12 and 0.16 respectively. Simulations indicate that  initial eccentricity scaling of elliptic flow is better observed in viscous fluid evolution than in ideal fluid evolution.
The result is important as it indirectly indicate that in Au+Au collisions, a low viscous  fluid is produced. Otherwise, in experiment more exact initial eccentricity scaling would have been observed. 

 \begin{figure}[t]
 \center\resizebox{0.35\textwidth}{!}{%
  \includegraphics{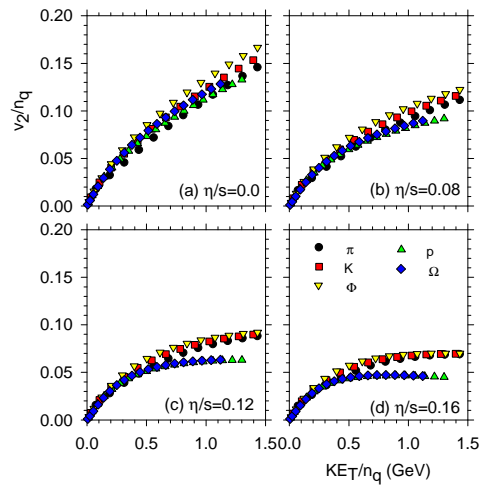} 
}
\caption{(color online) same as in Fig.\ref{F6} but as a function of constituent number scaled transverse kinetic energy $KE_T/n_q$.}
\label{F7}
 \end{figure} 
 
\subsection{Constituent quark number scaling}
While constituent quark number scaling is natural in a coalescence model, \cite{Molnar:2003ff,Scheibl:1998tk}, it is not indicated in a hydrodynamic model.
In a hydrodynamic approach, quark degrees of freedom are not manifestly evident. However, since in experiment, constituent quark number scaling is observed, and if hydrodynamics do faithfully represent the experimental data, one would expect, to be consistent with experiment, meson and baryon flow in the ratio 2:3 .

In Fig.\ref{F6}, elliptic flow for $\pi^-$, $K^+$, $p$, $\phi$ and $\Omega$, scaled by the constituent quarks number     in 20-30\% Au+Au collisions are shown as a function of   $p_T/n_q$. Quark number scaling of elliptic flow, as a function of $p_T/n_q$ naturally occur within a coalescence model \cite{Molnar:2003ff}. However, experimentally scaling is better observed as a function of transverse kinetic energy. In Fig.\ref{F7}, constituent quark number scaling as a function of $KE_T/n_q$ are shown. Hydrodynamics predictions for constituent quark number scaling, as a function of $p_T$ or $KE_T$ are similar, but the scaling seems to work better as a function of $KE_T/n_q$. 
At a fixed centrality, either as a function of $p_T$ or $KE_T$, constituent quark number scaling is only approximate in ideal fluid dynamics.  
In a viscous fluid, constituent quark number scaling seems to work better but separately for mesons and baryons. For fluid viscosity $\eta/s$=0.12-0.16, the
$n_q$-scaled $v_2$ of baryons and meson is approximately in the ratio 2:3 at $KE_T/n_q >$ 1 GeV. The result is expected. As shown in Fig.\ref{F3}, for $\eta/s$=0.12-0.16, elliptic flow for mesons and baryons are approximately same at large $p_T$. When scaled by the constituent quark number, naturally, the baryon and meson  flow are in the ratio 2:3. 

 \begin{figure}[t]
 \center
 \resizebox{0.35\textwidth}{!}{%
  \includegraphics{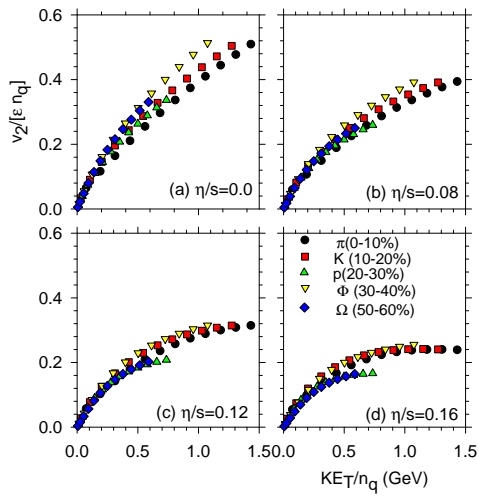} 
}
\caption{(color online) initial eccentricity and constituent quark number scaled  elliptic flow ($v_2/n_q\varepsilon$) for $\pi^-$, $K^+$, $p$, $\phi$ and $\Omega$ as a function of $KE_T/n_q$, in QGP fluid evolution with (a) $\eta/s$=0, (b) $\eta/s$=0.08, (c) $\eta/s$=0.12 and (d) $\eta/s$=0.16.} \label{F8}
\end{figure} 

In Fig.\ref{F8}, elliptic flow scaled by the initial eccentricity and constituent quark numbers as a function of transverse kinetic energy are shown. 
As before $\pi^-$, $K^+$, $p$, $\phi$ and $\Omega$'s are from 0-10\%, 10-20\%, 20-30\%, 30-40\%
and 50-60\% centrality Au+Au collisions. $\frac{v_2}{n_q\varepsilon}$ scaling is also an approximate in ideal fluid, or minimally viscous ($\eta/s$=0.08) fluid evolution. For example, at $KE_T/n_q \approx$ .6 GeV, $v_2/n_q\varepsilon$ of $\pi^-$ in 0-10\% collision and that of $\Omega$ in 50-60\% collision differ by $\sim$30\% in ideal and by $\sim$10\% in minimally viscous fluid evolution. 
At larger $KE_T$ the difference is even more. For higher values of viscosity $\eta/s$=0.12 and 0.16, at $KE_T/n_q >0.5$ GeV, $v_2/n_q\varepsilon$ for
mesons and baryons scale separately. The result is consistent with nearly perfect eccentricity scaling in viscous ($\eta/s$=0.12-0.16) fluid evolution (see Fig.\ref{F5}) and separate constituent quark number scaling for meson and baryons (see Fig.\ref{F7}). Violation of initial eccentricity and constituent quark number scaling in ideal or viscous hydrodynamics is also consistent with STAR and PHENIX experiments. Experimental data do not show the scaling.

\subsection{Universal scaling of elliptic flow}

Both the STAR  \cite{:2008ed} and PHENIX \cite{Taranenko:2007gb} collaboration observed universal scaling of elliptic flow. Elliptic flow scaled by the constituent quark number and the charged particle $v_2$, for different particle species    in different collision centrality   is approximately a single function of the quark number scaled transverse kinetic energy.  
The STAR collaboration observed the scaling for  $K_s^0$ $\Lambda$, $\Xi$ in 
0-10\%, 10-40\%, 40-80\% Au+Au collisions (see Fig. 12d in \cite{:2008ed}). 
In \cite{Taranenko:2007gb} PHENIX collaboration  showed that both in Au+Au and Cu+Cu collisions, the universal scaling is approximately valid for a large number of particle species in a variety of collision centrality  (see Fig.3 in  \cite{Taranenko:2007gb}).  

 \begin{figure}[t]
 \center
 \resizebox{0.35\textwidth}{!}{%
  \includegraphics{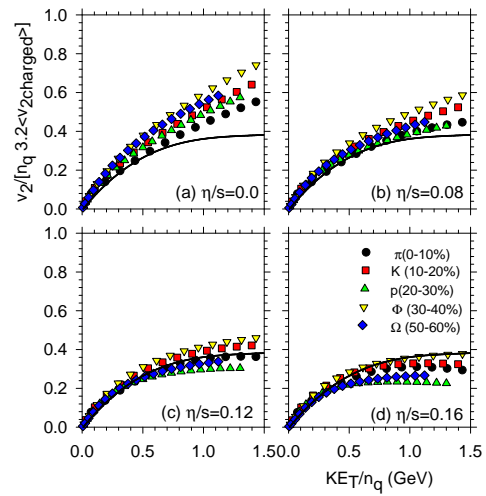} 
}
\caption{(color online) charged particle $v_2$ and constituent quark number scaled  elliptic flow ($v_2/n_q\varepsilon$) for $\pi^-$, $K^+$, $p$, $\phi$ and $\Omega$ as a function of $KE_T/n_q$, in QGP fluid evolution with (a) $\eta/s$=0, (b) $\eta/s$=0.08, (c) $\eta/s$=0.12 and (d) $\eta/s$=0.16. Approximate 'experimental' scaling function, as obtained in PHENIX experiment \cite{Taranenko:2007gb} is shown as the solid line.}
 \label{F9}
\end{figure}  

In Fig. \ref{F9}, we have examined the universal scaling in the hydrodynamic model. Hydrodynamical simulations for $v_2/(n_q*3.2*<v_2>_{ch})$, for $\pi$, $K$, $p$, $\phi$ and $\Omega$, as a function of $KE_T/n_q$ are shown. $<v_2>_{ch}$ 
is the appropriately weighted 
$p_T$ integrated flow for $\pi$, $K$ and $p$. In experiments, charged particles list is mainly populated by these three species. In Fig.\ref{F9}, the scaled flow, for $\pi$, $K$, $p$, $\phi$ and $\Omega$ are shown.    Contrary to experiments, in hydrodynamic simulations, irrespective of fluid viscosity, elliptic flow does not follow the   universal scaling. 

\section{Universal scaling violation and QGP viscosity} \label{sec4}

In previous sections we have studied several scaling properties of elliptic flow in ideal and viscous dynamics. It appears that ideal or viscous hydrodynamic simulations for elliptic flow in Au+Au collisions do not show the scaling properties exhibited in experiment. However, until now we have not compared hydrodynamic simulations for elliptic flow with experiments. 
In this section, we try to obtain an estimate of QGP  viscosity by confronting hydrodynamic simulations for elliptic flow with experiment.

\begin{table}[h]
\caption{\label{table2}  Departure of hydrodynamic simulations for elliptic flow of $\pi$, $K$, $p$, $\phi$ and $\Omega$, in the collision centrality 0-60\% from the PHENIX universal scaling function.}
\begin{ruledtabular} 
  \begin{tabular}{|c|c|c|c|c|c|}\hline
$\eta/s$ & $\pi$ & $K$ & $p$ & $\phi$ & $\Omega$ \\ \hline 
 0       & 0.87 & 2.20 & 3.04   & 5.36   & 6.17 \\ \hline     
0.08     & 0.25 & 0.67 & 0.76   & 1.53   & 1.62 \\ \hline
0.12     & 0.03 & 0.09 & 0.36   & 0.51   & 0.67 \\  \hline
0.16     & 0.05 & 0.09 & 0.91   & 0.97   & 1.50  \\ \hline  
 \end{tabular}\end{ruledtabular}  
\end{table}   
 
As stated earlier, PHENIX collaboration observed universal scaling of    elliptic flow for different particle species, in different centrality ranges of Au+Au and Cu+Cu collisions  \cite{Taranenko:2007gb}. 
In Fig.\ref{F9}, the solid line represents the approximate scaling function   as obtained in the PHENIX experiment \cite{Taranenko:2007gb}. Experimental scaling function does have fluctuations, more at large $KE_T/n_q$, which is presently ignored. Comparing the experimental scaling function with hydrodynamic simulations, we observe that, (i) in ideal fluid simulations, elliptic flow is over predicted for all the particle species, (ii) elliptic flow is also over predicted in minimally viscous fluid evolution, though the difference between theory and experiment is less than that in ideal fluid evolution, (iii) flow is under-estimated in evolution with $\eta/s$=0.16, and (iv) simulated flows show least departure from the experimental flow for $\eta/s$=0.12.

To measure the departure of simulated flow from the experimental 'universal scaling function',  we define a scaling violation function $F$,

\begin{equation}
F=\Sigma_i [(v_{2scaled}^{th})_i - v_{2scaled}^{ex}]^2
\end{equation}

\noindent where  $v_{2scaled}^{ex}$ is the approximate scaling function as obtained in PHENIX experiment (the solid line shown in Fig.\ref{F9}) and $(v_{2scaled}^{th})_i$ is the hydrodynamical simulations for the scaled flow $v_2/(n_q*3.2*<v_2>_{ch})$, in the $i-th$ centrality bin. As stated earlier, we have neglected resonance 
 contribution.   For low mass particles, present simulations for $v_2$ in low $p_T$, $p_T \leq$1 GeV, may not be reliable. 
Accordingly, we compute $F$ in the $p_T$ region $1 \leq p_T \le 3$ GeV. 
The quantity F is computed separately for particle species $\pi$, $K$, $p$, $\phi$ and $\Omega$, in 0-10\%, 10-20\%, 20-30\%, 30-40\%, 40-50\% and 50-60\% Au+Au collision. The results are shown in table.\ref{table2}. Departure from universal scaling function is mass dependent.
Hydrodynamic simulated flow for heavy mass particles deviate more from the universal scaling function than that of  a lighter particle. Scaling violation is also viscosity dependent. Scaling violation is more in ideal fluid evolution than in viscous fluid evolution. It may also be noted that for the viscosity to entropy ratio, $\eta/s$=0.12, the  violation is minimum. 

\begin{figure}[t]
 \center
 \resizebox{0.35\textwidth}{!}{%
  \includegraphics{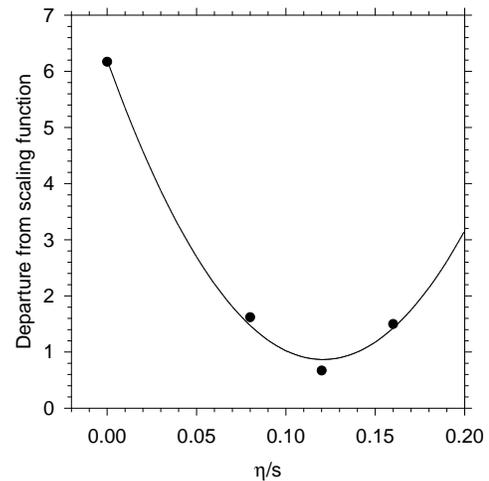} 
}
\caption{$\chi^2/N$ for the combined data set, $\pi^++\pi^-$, $K^++K^-$ and $p+{\bar p}$ as a function of viscosity to entropy ratio $\eta/s$. The solid line is a parabolic fit to the $\chi^2/N$ values.}
 \label{F10}
\end{figure} 

In Fig.\ref{F10}, the scaling violation function $F$, summed over all the particles ($\pi$, $K$, $p$, $\phi$ and $\Omega$) is shown as a function of $\eta/s$. The solid line in Fig.\ref{F10}
is a parabolic fit to $F$. From the minimum of the fit, we estimate viscosity to entropy ratio, for which simulated flows show least departure from the PHENIX experimental scaling function  as $\eta/s=0.12 \pm 0.03$.

The present estimate $\eta/s=0.12 \pm 0.03$ is well within the upper bound of viscosity obtained in \cite{Luzum:2008cw,Song:2008hj}. 
Within the error, the estimate also agrees with other estimates obtained in a hydrodynamical model  \cite{Drescher:2007cd,Lacey:2006bc}.   
 However, one must treat the phenomenological estimates with caution. As noted in \cite{Chaudhuri:2009uk}, systematic error in hydrodynamic evaluation of viscosity could be large. Indeed, in \cite{Chaudhuri:2009uk}, from a simultaneous fit to $\phi$ meson multiplicity, mean $p_T$ and integrated $v_2$, viscosity to entropy ratio was estimated as, $\eta/s=0.07 \pm 0.03 \pm 0.14$, where the 1st error is statistical and the 2nd one is the systematic error. Large ($\sim$ 200\%) systematic error arises due to uncertainty in initial time, initial energy density profile, initial fluid velocity, freeze-out condition, finite accuracy of computer code etc. Even then as noted  in  \cite{Chaudhuri:2009uk}  the source of systematic error is not exhaustive.
 
Before we summarise our results, few comments are in order. Present simulations indicate that  elliptic flow scaling is better observed in viscous hydrodynamics than in ideal hydrodynamics. The result is contradictory to our expectation. In viscous hydrodynamics, Cooper-Frye formula for invariant distribution ($\frac{dN}{dyd^2p_T}$), in addition to the thermal part,  includes a viscous correction  \cite{Chaudhuri:2008sj}. Since viscous correction introduces a microscopic scale, it is expected that   any scaling property that would have been observed in ideal fluid, will worsens. Simulation studies indicate that the contribution of the viscous part to elliptic flow can be large $\sim$ 50\% \cite{Molnar:2008xj,Song:2007fn}.  However, viscous contribution to elliptic flow does depend on various parameters, e.g.   if freeze-out temperature is lowered viscous correction decreases \cite{Chaudhuri:2008sj}. Viscous flux ($\pi^{\mu\nu}$) decreases rapidly with time and its contribution to Cooper-Frye decreases if fluid freezes out at later time or at lower temperature. Unlike in \cite{Chaudhuri:2008sj,Molnar:2008xj,Song:2007fn}, in the present simulations, a lattice based EOS with a confinement-deconfinement cross-over transition is used. There is no mixed phase and viscous fluxes decreases more rapidly than in 1st order phase transition. We have checked that in our simulations, at the late stage, fluid behaves nearly as an ideal fluid and viscous correction in Cooper-Frye is very small. Viscosity however changes the freeze-out surface. Changed freeze-out surface, but small viscous correction, may possibly be the reason for obtaining better scaling in viscous dynamics than in ideal dynamics. 

\section{summary}  \label{sec5}
To summarize, in ideal and viscous hydrodynamics, we have studied  scaling 
properties of elliptic flow in Au+Au collisions. Fluid was initialized to reproduce experimental $\phi$ meson multiplicity in 0-5\% centrality Au+Au collisions. Elliptic flow in 
 ideal or minimally viscous fluid evolution, show only approximate initial eccentricity scaling as a function of transverse kinetic energy. Initial eccentricity scaling is better observed in more viscous fluid evolution.  
Constituent quark number scaling is also an approximate scaling in ideal and minimally viscous dynamics. In more viscous fluid,   
 constituent quark number scaling seems of be obeyed separately  for mesons and  baryons. Elliptic flow in ideal or viscous dynamic also do not show experimentally observed 'universal scaling'. From a study of universal scaling violation in ideal and viscous evolution, we also obtain an estimate of QGP viscosity, $\eta/s=0.12 \pm 0.03$. However, the error does not include systematic error due to uncertain initial conditions in a hydrodynamic evolution. Systematic error in $\eta/s$ could be as large as $\sim$ 200\%.


\begin{thebibliography}{99}

\bibitem{:2008ed}
  B.~I.~Abelev {\it et al.}  [STAR Collaboration],
  Phys.\ Rev.\  C {\bf 77}, 054901 (2008).
\bibitem{Taranenko:2007gb}
  A.~Taranenko,
  J.\ Phys.\ G {\bf 34}, S1069 (2007)
  .

\bibitem{Molnar:2003ff}
  D.~Molnar and S.~A.~Voloshin,
  Phys.\ Rev.\ Lett.\  {\bf 91}, 092301 (2003)
.

\bibitem{Scheibl:1998tk}
  R.~Scheibl and U.~W.~Heinz,
  Phys.\ Rev.\  C {\bf 59}, 1585 (1999)
.
\bibitem{QGP3}
P.~F. Kolb and U. Heinz,
in {\it Quark-Gluon Plasma 3}, edited by R.~C. Hwa and 
X.-N. Wang (World Scientific, Singapore, 2004), p.~634.

\bibitem{Policastro:2001yc}
  G.~Policastro, D.~T.~Son and A.~O.~Starinets,
  Phys.\ Rev.\ Lett.\  {\bf 87}, 081601 (2001).
  
\bibitem{Arnold:2000dr}
  P.~Arnold, G.~D.~Moore and L.~G.~Yaffe,
  JHEP {\bf 0011}, 001 (2000),JHEP {\bf 0305}, 051 (2003).
  
\bibitem{Meyer:2007ic}
  H.~B.~Meyer,
  Phys.\ Rev.\  D {\bf 76}, 101701 (2007).
  
\bibitem{Nakamura:2005yf}
  A.~Nakamura and S.~Sakai,
  Nucl.\ Phys.\  A {\bf 774}, 775 (2006).
  
\bibitem{Gavin:2006xd}
  S.~Gavin and M.~Abdel-Aziz,
  Phys.\ Rev.\ Lett.\  {\bf 97}, 162302 (2006).
  
\bibitem{Drescher:2007cd}
  H.~J.~Drescher, A.~Dumitru, C.~Gombeaud and J.~Y.~Ollitrault,
  Phys.\ Rev.\  C {\bf 76}, 024905 (2007).
  
\bibitem{Lacey:2006bc}
  R.~A.~Lacey {\it et al.},
  Phys.\ Rev.\ Lett.\  {\bf 98}, 092301 (2007).
  
\bibitem{Adare:2006nq}
  A.~Adare {\it et al.}  [PHENIX Collaboration],
  Phys.\ Rev.\ Lett.\  {\bf 98}, 172301 (2007).

\bibitem{Luzum:2008cw}
  M.~Luzum and P.~Romatschke,
  Phys.\ Rev.\  C {\bf 78}, 034915 (2008).
  
\bibitem{Song:2008hj}
  H.~Song and U.~W.~Heinz,
  J.\ Phys.\ G {\bf 36}, 064033 (2009).
  
\bibitem{Chaudhuri:2009uk}
  A.~K.~Chaudhuri,
  Phys.\ Lett.\  B {\bf 681}, 418 (2009).

\bibitem{Chaudhuri:2009hj}
  A.~K.~Chaudhuri,
  arXiv:0910.0979 [nucl-th].
  
  \bibitem{IS79}
W. Israel, Ann. Phys. (N.Y.) {\bf 100}, 310 (1976);
W.~Israel and J.~M.~Stewart, Ann. Phys. (N.Y.) {\bf 118}, 349 (1979).

\bibitem{Heinz:2005bw}
  U.~W.~Heinz, H.~Song and A.~K.~Chaudhuri,
  Phys.\ Rev.\  C {\bf 73}, 034904 (2006).
  
\bibitem{Chaudhuri:2008sj} A.~K.~Chaudhuri,
 arXiv:0801.3180 [nucl-th].
 
\bibitem{Song:2008si}
  H.~Song and U.~W.~Heinz,
  Phys.\ Rev.\  C {\bf 78}, 024902 (2008).
   
\bibitem{lattice} 
Karsch F, Laermann E, Petreczky P, Stickan S and Wetzorke I, 
2001 {\it Proccedings of NIC Symposium} (Ed. H. Rollnik and D. Wolf, John 
von Neumann Institute for Computing, J\"ulich, NIC Series, vol.9, 
ISBN 3-00-009055-X, pp.173-82,2002.)

\bibitem{Cheng:2007jq}
  M.~Cheng {\it et al.},
  Phys.\ Rev.\  D {\bf 77}, 014511 (2008).

 \bibitem{note1} We have checked that with the lattice+HRG EOS, in ideal fluid dynamics, STAR measurements of $\frac{dN^\phi}{dy}$ and $<p_T^\phi>$ in 0-5\% Au+Au collisions are best explained with $T_F$=150 MeV.
 
\bibitem{Cooper:1974mv}
  F.~Cooper and G.~Frye,
  Phys.\ Rev.\  D {\bf 10}, 186 (1974).
 
\bibitem{Abelev:2007rw}
  B.~I.~Abelev {\it et al.}  [STAR Collaboration],
  Phys.\ Rev.\ Lett.\  {\bf 99}, 112301 (2007).
  
\bibitem{Voloshin:2007af}
  S.~A.~Voloshin  [STAR Collaboration],
  J.\ Phys.\ G {\bf 34}, S883 (2007) .

\bibitem{Voloshin:2006wi}
  S.~A.~Voloshin  [STAR Collaboration],
  AIP Conf.\ Proc.\  {\bf 870}, 691 (2006).
  
\bibitem{Bhalerao:2005mm}
  R.~S.~Bhalerao, J.~P.~Blaizot, N.~Borghini and J.~Y.~Ollitrault,
  Phys.\ Lett.\  B {\bf 627}, 49 (2005).
  
\bibitem{Hirano:2002ds}
  T.~Hirano and K.~Tsuda,
  Phys.\ Rev.\  C {\bf 66}, 054905 (2002).

\bibitem{Adare:2006ti}
  A.~Adare {\it et al.}  [PHENIX Collaboration],
  Phys.\ Rev.\ Lett.\  {\bf 98}, 162301 (2007).
  
\bibitem{Molnar:2008xj}
  D.~Molnar and P.~Huovinen,
  J.\ Phys.\ G {\bf 35}, 104125 (2008).

\bibitem{Song:2007fn}
  H.~Song and U.~W.~Heinz,
  Phys.\ Lett.\  B {\bf 658}, 279 (2008).
  
\end{thebibliography}
\end{document}